\documentclass[12pt]{article}

\usepackage{amssymb,amsmath}

\numberwithin{equation}{section}

\include{PhysNote}

\begin{document}

\begin{titlepage}
\begin{flushright}
CERN-TH-PH/2012-075
\end{flushright}
\vskip 1in
 
\begin{center}

{\Large {\bf CONFORMAL FIELD THEORIES WITH INFINITELY MANY CONSERVATION LAWS}
\footnote{Expanded version of a talk at the TH Journal 
Club on String Theory, CERN, February 27, 2012.}}

\vspace{4mm}

Ivan Todorov 

Theory Division, Department of Physics, CERN, CH-1211 Geneva 23,
Switzerland; e-mail: ivbortodorov@gmail.com 

and
 
Institute for Nuclear Research and Nuclear Energy\\
Tsarigradsko Chaussee 72, BG-1784 Sofia, Bulgaria\\
(permanent address); e-mail: todorov@inrne.bas.bg 

\end{center}

\begin{abstract}
 
Globally conformal invariant quantum field theories in a D-dimensional space-time (D even) have rational correlation functions and admit an infinite number of conserved (symmetric traceless) tensor currents. In a theory of a scalar field of dimension D-2 they were demonstrated  to be generated by bilocal normal products of free massless scalar fields with an O(N), U(N), or Sp(2N) (global) gauge symmetry [BNRT]. 

Recently, conformal field theories "with higher spin symmetry" were considered for D=3 in [MZ] where a similar result was obtained (exploiting earlier study of CFT correlators). We suggest that the proper generalization of the notion of a 2D chiral algebra to arbitrary (even or odd) dimension is precisely a CFT with an infinite series of conserved currents. We shall recast and complement (part of) the argument of Maldacena and Zhiboedov into the framework of our earlier work. We extend to $D=4$ the auxiliary Weyl-spinor formalism developed in [GPY] for $D=3$. The free field construction only follows for $D>3$ under additional assumptions about the operator product algebra. In particular, the problem of whether a rational CFT in 4D Minkowski space is necessarily trivial remains open.

\end{abstract}
\end{titlepage}
\vfill\eject
\section{Introduction}

\begin{flushright}
{\it Knowledge is recollection.} (attributed to Socrates) 
\end{flushright}
 
Higher dimensional conformal field theory (CFT) experienced a revival since the late 1990's. There were two independent developments.

One of them arose in an attempt to extend the notion of a 2D chiral algebra to higher dimensions. It led to the study of {\it globally conformal invariant} (GCI) quantum field theory (QFT) \cite{NT} and to the notion of a D-dimensional vertex algebra \cite{N}, \cite{BN}. Global conformal invariance led to a QFT with rational correlation functions that is only consistent for even space-time dimensions D. The resulting 4D CFT considered in \cite{NST02,NST03, NRT05, NRT08, BNRT} gives rise to bilocal fields whose operator product expansion (OPE) involves an infinite series of conserved higher spin (symmetric, traceless) tensor currents. Whenever such a bilocal field, say $W(x_1, x_2)$, is generated by products of scalar fields of dimension $D-2$,  it was proven to be a sum of normal products of free massless fields with a compact symmetry group.

The other, more popular development was an outgrow of the fashionable ``AdS/CFT correspondence'' (that brought, in the words of Sasha Polyakov, higher dimensional CFT to the masses). The particular branch of it which concerns us arose from the observation \cite{HS} that the free ${\mathcal N}=4$ superconformal Yang-Mills theory (with a vanishing 't Hooft coupling) on the boundary corresponds to a massless higher spin theory in the bulk. The systematic study of such higher spin theories beginning with \cite{FV} proceeded uninterrupted for some 25 years (see the recent partly expository papers \cite{V11, V12} where earlier references can be found) but only started attracting attention around 2002 \cite{M02, SS}, and especially after an exact duality was conjectured \cite{KP} between the (nonsupersymmetric!) 3D $O(N)$ vector model (in the free field limit) and Vasiliev's higher spin theory in $AdS_4$ (see \cite{DMR} for a later update and \cite{GGHR} for another point of view on these developments). The correlation functions in a 3D CFT (or in any odd dimensional CFT) are not rational. If one assumes, however, the presence of a higher spin conserved current, it is demonstrated in \cite{MZ} that the theory again involves an infinite number of conservation laws\footnote{A follow-up of this work \cite{MZ12} deals with a broken higher spin symmetry.}. This characteristic property thus provides a generalization of a chiral vertex algebra valid for both even and odd dimensions. The main result of \cite{MZ} is a proof of a statement similar to the one cited above: the presence of conserved currents of spin higher than two in a theory with a unique stress energy tensor implies for $D=3$ that the theory is generated by bilocal fields that appear as normal products of free massless fields.  This result can be viewed as an extension of the well known Coleman-Mandula theorem \cite{CM67} to the case of a CFT (in which there is no mass gap and no $S$-matrix). The triviality of the $S$-matrix is replaced by the identification  of the correlation functions of the stress-energy tensor with ones constructed  by free fields. (It was Gerhard Mack [M77] who advanced, back in the 1970's, the idea that the algebra of observables can be determined by the correlation functions (and the OPE) of the stress-energy tensor.) 

The aim of the present lecture is to display in some detail the relation between a central argument in (Secs. 5.3-5 of) 
[MZ] and our earlier work [BNRT]. Preparing the ground for a further study of four dimensional CFT with higher spin symmetry we extend to $D=4$ (in Sect. 3 and in A1-3) the auxiliary Weyl-spinor formalism developed in [GPY] for $D=3$. The rest of the paper presents a review and discussion of published results. Sect. 2 provides a brief survey (based on a single example) of our early work on conformal bilocal fields. We remark that the old argument remains valid without change for odd space-time dimensions. We begin Sect. 4 with a (rather superficial) review of part of the reasoning of Maldacena and Zhiboedov which brings us to the presence of a scalar field $J_0$ of dimension $D-2(=1$ for $D=3$) coupled to the stress energy tensor. We then demonstrate how the argument of [BNRT] (based on the study of unitary positive energy representations of infinite dimensional Lie algebras) allows to recover (and complement at one point) the result of [MZ]. In the discussion (Sect. 5) we reproduce the 4-point function of a conserved (spin one) current obtained very recently by Stanev [S] that does not appear to allow a free field construction.

\bigskip 

\section{Huygens locality; twist $D-2$ bilocal fields} 
\setcounter{equation}{0}

Finite conformal transformations can map spacelike into timelike separated pairs of points. Local commutativity of Wightman functions (\cite{SW}) and GCI then imply {\it Huygens' locality}:  commutators of observable (Bose) fields have support on the light cone,
\begin{equation}
\label{HL} 
(x_{12}^2)^n [\phi_1(x_1), \phi_2(x_2)] = 0 \,  (x_{12} = x_1 - x_2, \, x^2 = {\bf x}^2 - (x^0)^2)
\end{equation}
for n sufficiently large (if $\phi_i$ are fields of dimension $d_i$ and spin $s_i$ then it would suffice to take $n\geq d_i +s_i, i=1, 2$). Combined with energy positivity, Eq. (\ref{HL}) yields in turn rationality of correlation functions \cite{NT}.  These properties are only consistent with the known behaviour of free fields and conserved currents if the dimension $D$ of space-time is even. (The geometric reason for the difference between even and odd dimensions can be traced back to the fact that conformally compactified Minkowski space,
\begin{equation}
\label{Mcomp}
{\bar M} = ({\mathbb S}^{D-1}\times {\mathbb S}^1)/\{1, -1\},
\end{equation}
is only orientable for $D$ even. In particular, the real projective plane, that is isomorphic to the 2-sphere with identified opposite points - the first factor in (\ref{Mcomp}) for $D=3$ - is a prime 
example of a non-orientable manifold.)  

The spectacular development of 2D CFT in the 1980's was based on the preceding 
progress in the study of infinite dimensional (Kac-Moody and Virasoro) Lie 
algebras and their representations. This tool did not seem to work in higher dimensions, however. It was proven 
\cite{B} that scalar Lie fields do not exist in three or more dimensions. It is all the more interesting that this no-go result does not apply to {\it bilocal} fields, in particular to the field $W(x_1, x_2)$ defined as the contribution of twist $D-2$  currents to the operator product expansion (OPE) of a pair of scalar fields of dimension $D-2$:
\begin{equation}
\label{W}
(\Delta_{12}^+)^{-1}\phi^*(x_1) \phi(x_2) = N \Delta_{12}^+ + W(x_1, x_2) + 
O(x_{12}^2), \, \, W(x_1, x_2)^* = W(x_2, x_1).
\end{equation}
If the local field is real, $\phi = \phi^*$, then the bilocal field is 
symmetric, $W(x_1, x_2) = W(x_2, x_1)$; the positive constant $N(=N_\phi)$ gives
 the normalization of the 2-point function of $\phi$:
\begin{equation}
\label{2ptf}
<\phi(x_1)\phi(x_2)^*> = N (\Delta_{12}^+)^2,
\end{equation}
while $\Delta_{ij}^+ = \Delta^+(x_i - x_j)$ is the Wightman 2-point function of 
a free massless scalar field $\varphi$, in D space-time dimension:
\begin{equation}
\label{Delta}
\Delta_{12}^+ = \frac{\rho_{12}^{2-D}}{(D-2)|{\mathbb S}^{D-1}|}, \,
\rho_{12} = (x_{12}^2 + i0x_{12}^0)^{\frac{1}{2}}, \, |{\mathbb S}^{D-1}| = \frac{2\pi^{D/2}}{\Gamma(D/2)}.
\end{equation}
 It is a consequence of the theory 
of the unitary positive energy irreducible representations of the (quantum mechanical) conformal group $Spin(D, 2)$ (see \cite{M} for $D=4$ and \cite{FF} for arbitrary $D$) that symmetric tensor currents of twist $D-2$ (in a QFT satisfying Wightman positivity) are conserved. As $W$ has by definition an expansion in terms (of suitable integrals) of such currents (see \cite{FGG}, \cite{DMPPT}, \cite{NST02}), it then follows \cite{NST03} that $W$ is {\it biharmonic} (i.e. satisfies the d'Alembert equation in each argument). Global conformal invariance and rationality allow to compute the 4-point function
\begin{eqnarray}
\label{4pt}
<W(1, 2)W(3, 4)> = N \Delta_{14}^+ \Delta_{23}^+ = \sum_{s=0}^\infty <W(1, 2)\Pi_s W(3, 4)>, \, \\ \nonumber
(W(i, j) \equiv W(x_i, x_j))
\end{eqnarray}
(provided that $\rho_{12}^{D-2}\phi(x_1)\phi(x_2)\rightarrow 0$ for $x_1\rightarrow x_2$). Here $\Pi_s$ is the (orthogonal) projection operator to the ``spin $s$ subspace'' - more precisely to the subspace generated by (neutral) scalar (for $s=0$), vector (for s=1) and rank $s$ symmetric traceless tensors
of scale dimension $d_s = D-2+s$. We note that each {\it conformal partial wave} (cf. \cite{DMPPT}, \cite{DO}) - i.e., each term in the right hand side of (\ref{4pt}) - is not a rational function (it involves logarithms); only the infinite sum gives the rational expression in the first equation (\ref{4pt}). Thus, GCI implies the presence of an infinite set of (twist $D-2$) conserved currents. Furthermore, Eq. (\ref{W}), the first equation  (\ref{4pt}) and the analysis of the GCI 4-point function $<\phi^*(x_1)\phi(x_2)\phi^*(x_3)\phi(x_4)>$ imply the commutation relations
\begin{eqnarray}
\label{CRW}
[W(1,2), W(3,4)] = \Delta_{23} W(1,4) + \Delta_{14} W(3,2) +N\Delta_{12,34}, \, \\ \nonumber 
\Delta_{ij} := \Delta_{ij}^+ - \Delta_{ji}^+, \, \Delta_{12,34}= \Delta_{14}^+\Delta_{23}^+ - \Delta_{41}^+\Delta_{32}^+.
\end{eqnarray} 
We note that Eq. (\ref{CRW}) also makes sense in odd space-time dimensions when the (pure imaginary) commutator function $\Delta_{ij}$ violates Huygens' locality (it is an integrable function with support in the closure of the cone of time-like separations: $(x_{ij}^0)^2 -{\bf x}_{ij}^2 (= -x_{ij}^2) \geq 0$). In all cases the bilocal field W and the unit operator span an infinite dimensional Lie algebra (that was identified as a central extension of $u(\infty, \infty)$ or of $Sp(2\infty, {\mathbb R})$  for $\phi = \phi^*$). It was demonstrated (first under somewhat restrictive assumptions in \cite{NST02} then, gradually in full generality in \cite{NRT05} and \cite{BNRT}) that for unitary positive energy representations of the arising infinite dimensional Lie algebras the parameter N in (\ref{CRW}) (and in similar formulas for more general bilocal fields considered in these papers) has to be a (positive) integer. This implies the existence of a free field construction of the bilocal field with the above properties. We shall expand on this statement in Sect. 4  after displaying its relevance to the argument of \cite{MZ}.

\bigskip

\section{Two and three point spin-tensor invariants}
\setcounter{equation}{0}

The study of conformally invariant 3-point functions of conserved (vector and tensor) currents has started over half a century ago \cite{S71} \cite{MS}. An early result was the construction \cite{S88} of the 3-point functions  of a $U(1)$ current $J_1$ and of the stress-energy tensor $J_2$ for $D=4$ ending with the observation that all the appearing structures are reproduced by composites of free scalar, spinor and Maxwell fields. This was followed (six years later) by a study of conformal invariant 3-point functions for arbitrary $D$ in \cite{OP} and (after five more years) in \cite{AF}. Lately (after another ten years) generating functions for all 3-point functions of three currents $J_s$ of arbitrary spins $s$ in $D=3$ were written down \cite{GPY}; a set of algebraic generating functions for 3-point correlators in  $D=4$ were proposed in \cite{S}; the result is been extended to arbitrary $D$ in \cite{Z}. Methods - old and new - of 
constructing conformal invariant correlation functions and OPE continue to 
attract attention - see e.g. \cite{CPPR, S-D},

One device (which goes back to Weyl when applied to the unitary group and to 
Penrose\footnote{In fact, the ideas associated nowadays with ``Penrose twistor 
theory'' go back to Julius Pl\"ucker (1801-1868) who introduced a parametrization of 3D projective lines in terms of a quadric in the projective space ${\mathbb P}^5$ (published posthumously, in Pl\"ucker's memoir ``New geometry of space based on the consideration of a line as a space element'' prepared by his assistant Felix Klein and by Alfred Clebsch in 1868-69 - see the lively historical survey \cite{G}). I thank Raymond Stora for reminding me of Pl\"ucker's pioneering role.} (and Witten) when applied to the conformal group), used systematically in \cite{GPY} for $D=3$, consists in replacing spin-tensor indices by 2-component spinors. We shall work it out here and in the Appendix (see A1-3) for the $D=4$ case and will indicate the reduction to $D=3$ in Sect. 4 (and in A4). 

To set the stage we shall review some basic facts on the unitary positive energy irreducible representations (UPEIR) of the 4-dimensional quantum mechanical conformal group ${\mathcal C} = SU(2, 2)$ \cite{M} (for reviews and further developments - see \cite{TMP, T, FF}). Local field UPEIR are induced by finite dimensional irreducible representations (IR) of the 11-parameter parabolic subgroup ${\mathcal P}$ of ${\mathcal C}$ that leaves the point $x = 0$ invariant,
\begin{equation}
\label{PNc}
{\mathcal P} = N_c \rtimes ({\mathbb R}_+ \times SL(2,{\mathbb C})), 
\end{equation} 
where ${\mathbb R}_+$ is the (multiplicative) dilation group (of positive reals), $N_c$ is the 4-parameter (nilpotent) abelian group of special conformal transformations 
\begin{equation}
n_c: x \rightarrow \frac{x + c x^2}{1 + 2cx + c^2 x^2}
\end{equation}
 (that acts trivially on $x=0$). The inducing representations are labeled by the $SL(2, {\mathbb C})$ weight $(s_1, s_2), s_i = 0, \frac{1}{2}, 1, ...$ and the
scale dimension $d$ that labels the IR of the 1-parameter dilation group; $2s_1 (2s_2)$ count the udotted (dotted) spinor indices $A, {\dot B} = 1, 2$. We shall contract the indices $(A, {\dot B})$ of local fields by complex 2-component spinors $\lambda = (\lambda_A)$
(and their conjugates). Let ${\tilde \Lambda}\in SL(2, {\mathbb C})$ and $\Lambda = 
\Lambda(\pm {\tilde \Lambda})$ be the corresponding (proper) Lorentz transformation; let further 
$a\in {\mathbb R}^4$ parametrize coordinate translation and let $\rho > 0$. Let, finally, $\phi =
\phi(x; \lambda, {\bar \lambda})$ be a local field of Lorentz weight $(s_1, s_2)$ and dimension $d$. The transformation law of $\phi$ under Poincar\'e transformations and dilation (the subgroup of ${\mathcal C}$ conjugate to ${\mathcal P}$ that leaves invariant the tip of the cone at infinity) can be written as:
\begin{eqnarray}
\label{P'}
U({\tilde \Lambda}, a) \phi(x; \lambda, {\bar \lambda})U({\tilde \Lambda}, a)^{-1} =
\phi(\Lambda x +a; \lambda {\tilde \Lambda}^{-1}, ({\tilde \Lambda}^*)^{-1} {\bar \lambda}),  \nonumber \\
U(\rho)\phi(x; \lambda, {\bar \lambda})U(\rho)^{-1} = \rho^t \phi(\rho x; \rho^{1/2}\lambda, 
\rho^{1/2}{\bar \lambda}) = \rho^d \phi(\rho x; \lambda, {\bar \lambda}),  \nonumber \\
d = t + s_1 + s_2.
\end{eqnarray}      
Here $t$ is the {\it twist} (a terminology usually applied to the case $s_1 = s_2 = r/2$
where $r$ is the {\it spin} of the (rank $r$) symmetric tensor current). The special conformal transformation law of $\phi$ can be deduced from the knowledge of the action of the Weyl 
inversion (see Appendix A2). We shall write down the corresponding infinitesimal conformal law:
\begin{eqnarray}
\label{C}
[C_\mu, \phi(x; \lambda, {\bar \lambda})] = [x^2 \partial_\mu - 2x_\mu (x\partial + t) - (\lambda
{\tilde \sigma}_\mu x^\nu \sigma_\nu \frac{\partial}{\partial \lambda} + c.c.)]
\phi(x; \lambda, {\bar \lambda}) \nonumber \\   = [x^2 \partial_\mu - 2x_\mu (x\partial + d) - (\lambda \sigma_{\mu \nu} x^\nu\frac{\partial}{\partial \lambda} + c.c.)]\phi(x; \lambda, {\bar \lambda}) \, \,
\end{eqnarray}  
(c.c. standing for complex conjugation). The Pauli matrices $\sigma^\mu$ and ${\tilde \sigma}_\mu$ differ just by their transformation properties, which are so chosen that $\zeta_\mu = \lambda {\tilde \sigma}_\mu {\bar \lambda}$ behaves as a (null) 4-vector while $\frac{\partial}{\partial{\bar \lambda}}\sigma^\mu\frac{\partial}{\partial x^\mu}\frac{\partial}{\partial \lambda}$ is a Lorentz invariant differential operator (see Eq. (\ref{eqs}) and Appendix A). The $sl(2,{\mathbb C})$ generators $\sigma_{\mu \nu}$ (\ref{smn}) obey the self duality relations obtained from $\sigma_{12} = i\sigma_{03} = i\sigma_3$ by cyclic permutations of the subscripts $(1, 2, 3)$. 

For a pair of points $x_1, x_2$ in 4D Minkowski space and a pair of spinors $\lambda_1, {\bar \lambda}_2$ (of undotted and dotted indices, respectively) we make correspond a (complex) conformal invariant (see Appendix A2)
\begin{equation}
\label{P12} 
P_{12} = \lambda_1 {\check x}_{12}{\tilde \sigma} {\bar \lambda}_2, \, {\check x}{\tilde \sigma}:= \frac{{\tilde x}}{x^2}, \, 
{\tilde x} = x^0 + x^1 \sigma_1 + x^2\sigma_2 + x^3\sigma_3.
\end{equation}
(For a reminder of the Pauli matrix gymnastics and of the conformal properties of $\lambda_i$ we again refer to the Appendix.) 
In particular, the 2-point function of the Weyl spinor field $\psi(x, \lambda)$ and its conjugate has the form
\begin{equation}
\label{psi}
w_\psi:=<\psi(x_1,\lambda_1)\psi^*(x_2,{\bar \lambda}_2)> = \frac{P_{12}}{2\pi^2 \rho_{12}^2}
\end{equation}
where $\rho_{12}$ and $P_{12}$ are given by (\ref{Delta}) and (\ref{P12}). 
Besides the 2-point invariants $P_{ij} = - {\bar P}_{ji}$ one also has real three 3-point invariants $L_i, i=1, 2, 3$ (denoted by $-Q_i$ for $D=3$ in \cite{GPY}) obtained from anyone of them by cyclic permutations:
\begin{eqnarray}
\label{Li}
L_1(=L^1_{23})= \lambda_1({\check x}_{12} + {\check x}_{31}){\tilde \sigma}{\bar 
\lambda}_1 =: \zeta_1 ({\check x}_{12} - {\check x}_{13}),  \nonumber \\
L_2 (=L^2_{31})= \zeta_2 ({\check x}_{23} + {\check x}_{12}), \, L_3 (=L^3_{12})= 
\zeta_3 ({\check x}_{23} - {\check x}_{13}), \, \zeta_i^2 = 0.
\end{eqnarray}
The lightlike ``polarization'' 4-vectors $\zeta_i$ given by the sesquilinear expressions 
$\zeta_{i\mu} = \lambda_i \sigma_\mu {\bar \lambda}_i$ also appear in the product 
\begin{equation}
\label{PP}
2P_{12} P_{21} = \zeta_1 \zeta_2 {\check x}_{12}^2 - 2({\check x}_{12}\zeta_1)({\check x}_{12}\zeta_2)=: R_{12}, 
\end{equation}
a standard expression, encountered in the conformal invariant 2-point functions of rank $r$ symmetric tensor currents $J_r(i)=J_r(x_i, \zeta_i)$:
\begin{equation}
\label{2JDs}
<J_r(1)J_r(2)>= C(r, D)\frac{(R_{12})^r}{(\rho_{12}^2)^{D-2}}.
\end{equation}
The conformal 2-point functions of twist $D-2$ currents automatically satisfy the conservation law which can be written as follows in terms of the lightlike vector $\zeta$:
\begin{equation}
\label{zet*}
\zeta^* \partial_x J_r(x, \zeta) =0 \, \, \mbox{where} \, \, \zeta^* =(\zeta\partial_\zeta +\frac{D}{2} -1)\partial_\zeta -\frac{1}{2}\zeta \partial_\zeta^2. 
\end{equation}
 (The systematics of replacing vector indices by lightlike vectors has been spelled out in \cite{BT} where the above $\zeta^*$ is defined as the {\it interior derivative} on the lightcone. The term with the d'Alembert operator $\partial_{\zeta_1}^2$ does not contribute when acting on $(R_{12})^r$ since $r(r-1)(x^2\zeta_2 -2(x\zeta_2)x)^2 = 0$ for $\zeta_2^2=0$.)  We note that even for integer spin skew-symmetric tensor fields it is convenient to use the spinorial invariants $P_{12}$ rather than the quadratic combination (\ref{PP}). For instance the 2-point function of the selfdual Maxwell tensor 
\begin{equation}
\label{Fomeg}
F(x, \lambda):=\frac{1}{2}F^{\mu \nu}(x)\omega_{\mu \nu}, \, \omega_{\mu \nu}=\lambda 
\sigma_{\mu \nu}\epsilon \lambda \, \, (\omega^{\mu \rho}\omega_{\rho \nu} = 0)
\end{equation} 
and its hermitean conjugate, and their 3-point function with the (electromagnetic) stress-energy tensor $T(=T_F)$ have the form
\begin{eqnarray}
\label{F}
<F(x_1, \lambda_1) F^*(x_2, {\bar \lambda}_2)>(\equiv <F(1)F(2)^*>) \sim \frac{P_{12}^2}{\rho_{12}^2}, \, \nonumber \\ <F(1)F(2)^* T(3)>\sim\frac{P_{13}^2 P_{32}^2}{\rho_{13}^2\rho_{23}^2}.  
\end{eqnarray}
The free field equations for $\phi = \psi, F$ and the conservation law for $J$, consistent with the above 2- and 3-point functions, can be written as:
\begin{equation}
\label{eqs}
 \sigma^\mu \partial_\mu \frac{\partial}{\partial \lambda} \phi(x; \lambda)= 0, \, \, \partial_\mu = \frac{\partial}{\partial x^\mu},
\, \frac{\partial}{\partial {\bar\lambda}}\sigma^\mu \partial_\mu \frac{\partial}{\partial \lambda} J_r(x; \lambda, {\bar \lambda}) = 0;
\end{equation}
Here $\phi(x, \lambda)$ is a free massless field transforming under the (complex) representation $(s, 0)$ of the spinorial Lorentz group $SL(2,{\mathbb C})$ (that is a homogeneous polynomial in $\lambda$ of degree $2s$). The real part of the cyclic n-point invariant
\begin{equation} 
\label{P1n}
2Re(P_{12}P_{23}...P_{n-1 n}P_{n1}) = P_{12}P_{23}...P_{n1}+(-1)^n P_{1n}P_{n n-1}...
P_{21} 
\end{equation}
can be expressed in principle in terms of $R_{ij}$ (\ref{PP}) and $L_k$ 
(\ref{Li}). The result is relatively simple for the 3-point function 
(derived in Appendix A2):
\begin{equation}
\label{PPP}
2P_{12} P_{23} P_{31} - 2P_{21} P_{32} P_{13}= R_{12} 
L_3 + R_{23} L_1 + R_{13} L_2 + 2 L_1 L_2 L_3. \,
\end{equation}
It becomes much more involved (and less satisfactory) for higher point 
correlation functions (see Appendix A3). The expression (\ref{PPP}) appears in
the numerator of the 3-point function of a non-abelian current $J^a = 
:\psi^*(x, {\bar \lambda})t^a\psi(x, \lambda):$ where the free Weyl field 
$\psi$ is equipped with internal symmetry indices coupled to the matrices $t^a$
 that span a finite dimensional representation of the associated Lie algebra. 
The 3-point function of the $U(1)$-current (corresponding to $t^a=1$), on the 
other hand, is proportional instead to the imaginary part of the triple product,
 $2 i Im(P_{12}P_{23}P_{31})=P_{12}P_{23}P_{31} + P_{21}P_{32}P_{13}$, which is the 
odd structure (involving the Levi-Civita tensor), computed in \cite{S}:
\begin{eqnarray}
\label{Podd}
2i(P_{12}P_{23}P_{31}+P_{13}P_{32}P_{21}) = \zeta_1\wedge\zeta_2\wedge\zeta_3\wedge
({\check x}_{13}^2({\check x}_{23}^2 x_{12}+{\check x}_{12}^2 x_{23}) -{\check x}_{12}^2
{\check x}_{23}^2 x_{13}) \, \, \, \, \, \nonumber \\
+2{\check x}_{23}^2 \, \zeta_1(x_{12} + x_{13}) \, \zeta_2\wedge\zeta_3 \wedge{\check x}_{12}\wedge
{\check x}_{13}+2{\check x}_{13}^2 \zeta_2(x_{12}-x_{23}) \zeta_1\wedge\zeta_3 \wedge 
{\check x}_{12}\wedge {\check x}_{23} - \, \, \, \nonumber \\
2{\check x}_{12}^2 \zeta_3(x_{13}+x_{23})\zeta_1\wedge \zeta_2 \wedge 
{\check x}_{13} \wedge {\check x}_{23}, \, \, \, \, \, \,
\end{eqnarray} 
where $x\wedge y\wedge z\wedge w := \epsilon^{\alpha \beta \mu \nu} x_\alpha y_\beta
z_\mu w_\nu$. The 3-point function of the rank two current that coincides with 
the fermionic stress-energy tensor $J_2 = T_\psi$, however, admits a factor 
given by the "parity even" product 
(\ref{PPP}):
\begin{eqnarray}
\label{J2f}
<J_2(1)J_2(2)J_2(3)> = \frac{8}{\pi^6}\frac{P_{12}P_{23}P_{31} - P_{13}P_{32}P_{21}}{\rho_{12}^2
\rho_{23}^2 \rho_{13}^2}[3(R_{12}L_3 + R_{23}L_1  \nonumber \\ 
+ R_{13}L_2)  - 2(P_{12}P_{23}P_{31} - P_{13}P_{32}P_{21})]. \,  \, \,
\end{eqnarray} 

The identity (\ref{PPP}) also allows to simplify the expression for the 3-point function of the stress-energy tensor $T(x, \zeta)$ for the Maxwell field (first computed in \cite{S88}), 
\begin{eqnarray}
\label{TF}
<T(1)T(2)T(3)>=\frac{16}{\pi^6 \rho_{12}^2\rho_{23}^2\rho_{13}^2}[R_{12}R_{23}R_{13} +  \nonumber   \\
(R_{12}L_3+R_{23}L_1+R_{13}L_2 + 2L_1L_2L_3)^2 ]=\frac{64(P_{12}^2P_{23}^2P_{31}^2+P_{21}^2P_{32}^2P_{13}^2)}
{\pi^6 \rho_{12}^2\rho_{23}^2\rho_{13}^2}. 
\end{eqnarray}
More generally, let $\phi(x, \lambda)$ be a {\it chiral conformal field} of twist one that transforms under the representation $(s, 0)$ of $SL(2,{\mathbb C})$. Then its 2-point 
function is given by   
\begin{equation}
\label{2s}
<\phi(x_1, \lambda_1) \phi^* (x_2, {\bar \lambda}_2)> = N_s \frac{P_{12}^{2s}}{\rho_{12}^2}
\end{equation} 
and satisfies the free field equation (\ref{eqs}). As demonstrated in \cite{S} for each $r\geq 2s$ there exists a conserved current given by a sesquilinear combination of (derivatives of) $\phi, \phi^*$. In particular, the rank $r=2s$ tensor current $J_r(x; \lambda, {\bar \lambda}) = 
:\phi^* (x, {\bar \lambda})\phi(x, \lambda):$ has a 3-point function of the form
\begin{equation}
\label{3s}
<J_r(1)J_r(2)J_r(3)>= C_r\frac{(P_{12}P_{23}P_{31})^r + (P_{13}P_{32}P_{21})^r}{(\rho_{12}\rho_{23}\rho_{13})^2},
\end{equation}
where $J_r(i) =J_r(x_i; \lambda_i,{\bar \lambda}_i)$. (For $r>2s$ this 3-point function contains the factor $(P_{12}P_{23}P_{31})^{2s} + (- P_{13}P_{32}P_{21})^{2s}$.) One has, in fact, the following generalization of the Weinberg-Witten theorem \cite{WW}: a conserved rank r current $J_r$ can couple to a spin s chiral field $\phi$ (in the sense that the 3-point function $<\phi(1)\phi(2)^* J_r(3)>\neq 0$) if and only if $2s\leq r$. Curiously, conformal invariance does not forbid such a coupling for higher point correlation functions or composite fields. An interesting example of this type has been constructed in \cite{S}. If $J(x, \zeta)(=J_1)$ is a conserved $U(1)$- current and $F(x,\lambda)$ has the property of the selfdual Maxwell field (with 2- and 3-point functions given by (\ref{F})) then, in accord with the Weinberg-Witten theorem,  there is no non-trivial conformal 3-point function $<F(1)F(2)^*J(3)>$ (consistent with current conservation) but the product of two $J$'s may be coupled to $F^2$. Indeed, using the new 3-point invariants (\ref{RoL}) that involve the isotropic selfdual skew symmetric tensor $\omega_{\mu \nu}$ (\ref{Fomeg}) (see Appendix A2) one can write
\begin{equation}
\label{JJF}
<J(1)J(2)F^2(3)> \sim \frac{RL_{13} RL_{23}}{\rho_{12}^4},
\end{equation}  
a particularly interesting example since it is not reproduced by free fields.
\bigskip

\section{Reduction to $D=3$. Infinite Lie algebras}
\setcounter{equation}{0}
 
The passage from four to three dimension in the above expressions amounts to 
reducing the complex transverse variable $y=x_1+ix_2$ to a real one, 
$y ={\bar y}$ and regarding $\lambda_i$ as {\it real} spinors (the 4D spinorial
 Lorentz group $SL(2, {\mathbb C})$ going to $SL(2,{\mathbb R})$ in 3D). Introducing also the 
lightcone variables $x^\pm = x^0\pm x^3$ we can present the two-by-two real 
symmetric matrix ${\check x{\tilde \sigma}}$ in the form:
\begin{equation}
\label{x3D}
{\check x}\sigma = \frac{{\tilde x}}{x^2}, \, \, {\tilde x} = x^+\sigma_+ + 
x^-\sigma_-+ y\sigma_1, \, \mbox{where} \,\sigma_\pm =\frac{1}{2}(1\pm \sigma_3).
\end{equation}
Their conformal transformation properties are described in Appendix A4. 
The six 2- and 3-point invariants $P_{ij}, L_k, \, i \neq j, k = 1, 2, 3$ are all real and continue to obey the cubic relation (\ref{PPP}) (while $P_{12}P_{23}P_{31}+P_{13}P_{32}P_{21}=0$).
Any polynomial of $\{P_{ij}, L_k\}$ that is homogeneous of degree $2s_i$ with respect to $\lambda_i, \lambda_i'$ when multiplied by 
$(\rho_{12} \rho_{23} \rho_{13})^{-1}$ gives a tensor structure for $<J_{s_1} J_{s_2} J_{s_3}>$ consistent with conformal invariance. The P's and L's are the building blocks of the generating functions of 3-point functions of higher spin currents. To write down the most general conformal 3-point function one has to take into account also parity violating structures computed in [GPY]
(which involve the rank three Levi-Civita tensor). Remarkably, all the basic 3-point functions become rational upon multiplying by $\rho_{12}\rho_{23}\rho_{13}$.

Current conservation is not automatic since the general conformal 3-point functions need not belong to a unitary theory. When imposed, the parity violating structures $<J_{s_1}J_{s_2}J_{s_3}>_{odd}$ are found to only exist if $s_i$ satisfy the triangle inequality, $|s_1-s_2|\leq s_3 \leq s_1 + s_2$ and then to be unique (for each triple of such $s$ - see Sect. 3 of \cite{GPY}). The ``parity preserving'' 3-point functions, on the other hand, are just the two Bose and Fermi structures coming from products of derivatives of free fields.  A basic assertion of \cite{GPY} (for which an improved argument is presented in Sect. 6.5 of \cite{MZ}) says that the presence of higher spin conserved currents (with $s\geq 4)$) excludes the non free ``odd structures'' for $D=3$ . 
\smallskip

Here are some of the steps in the argument of Maldacena and Zhiboedov. 

Let $J_s$ be a rank $s(>2)$ conserved tensor current. Writing $j_s$ for its contraction with a fixed a lightlike vector $\zeta$ such that 
\begin{equation}
\label{x+}
x\zeta = \lambda {\tilde x}\lambda = x^+ \,  \, (\lambda = (1, 0)),  \, j_s(x) := J_s(x, \zeta)
\end{equation}
we assume that the charge (of scale dimension $s-1$)
\begin{equation}
\label{Qs}
Q_s := \int_{x^+=const} j_s dy dx^-
\end{equation}
is a well defined operator (that annihilates the vacuum). The dimensionless charge $Q_1$ is central while for $s>1$ the commutators $[Q_s, j_{s'}]$ have the form
\begin{equation}
\label{Qj}
[Q_s, j_{s'}] = \sum_{s''=max(s'-s+1, 0)}^{s'+s-1} a_{ss's''}\partial^{s'+s-1-s''} j_{s''}, \, \partial \equiv \frac{\partial}{\partial x^-}
\end{equation}
such that the nonvanishing of $a_{slm}$ is equivalent to the nonvanishing of $a_{sml}$. One proves - taking into account that (integrals of) $J_2$ 
generate all conformal transofrmations - that $[Q_s, j_2]$ involves (for $s>1$) a non-zero term proportional to $\partial j_s$ (in particular, $Q_2$ is just the component $P_-$ of the energy momentum operator, i.e. the generator of translation in the variable $x^-$). 

Using the assumed uniqueness of the field $J_2$, which coincides with the stress-energy tensor, as well as the fact that the OPE of two fermionic currents $(j_2 j_2)_f$, has a different singularity than that of two bosonic ones, $(j_2 j_2)_b$, one concludes that if $<j_2j_2j_2>_b\neq 0$ then $<j_2j_2j_2>_f=0$ and vice versa. In the first case, exploiting the presence of a conserved charge $Q_4$, one proves that there exists a scalar field $j_0$ (of dimension 1) such that $<(j_2j_2)_b j_0>\neq 0$. Furthermore, the OPE of a pair of $j_0$'s gives rise to a (hermitean) bilocal field  which contains (much like our $W(x_1, x_2)$ of Sect. 2) the stress-energy tensor and the higher spin conserved currents in its OPE (see below). Similarly, in the second case, one proves the existence of a scalar field of dimension two that behaves as a (skewsymmetric) product of two (2-component free massless) Majorana spinor fields
(in $D=3$). Fixing attention to the first (bosonic) case we shall demonstrate that the remaining argument (contained in Secs. 5.3-5 of \cite{MZ}) can be reduced to a special case of the results of \cite{BNRT} which
we proceed to summarize, noting that they remain valid for odd space-time dimensions. 

The OPE of twist two scalar fields like $J_0$ gives rise to a family of bilocal 
scalar fields $\{ V_M(x_1, x_2)\}$ where the $M$'s span a real matrix algebra 
closed under transposition. (The field $W(x_1, x_2)$ of Eqs. (\ref{W}) 
(\ref{CRW}) can be viewed as a special case in 
which the possible values of $M$ are either the (two-by-two) unit matrix or the
 skew-symmetric matrix $\epsilon =i\sigma_2$.) One proves (in Secs. 2 and 3 of 
\cite{BNRT} (b)) that it generates an infinite Lie algebra of one of the 
following three types\footnote{They correspond to irreducible real matrix 
algebras, whose commutants, according to the real version of Schur's lemma, are
 isomorphic to ${\mathbb R}, {\mathbb C}$ and ${\mathbb H}$ (the quaternions), 
respectively. (The last possibility yielding a $USp(2N)$ gauge symmetry is missed in [MZ].)}: a central extension of (i) the real symplectic algebra 
$sp(2\infty, {\mathbb R})$, (ii) the pseudounitary algebra $u(\infty, \infty)$,
 and (iii) the non-compact orthogonal algebra  $so^*(4\infty)$. Furthermore, one
 proves that in each case the unitary positive energy representations of the 
corresponding infinite dimensional Lie algebra are generated by a finite set of
 free massless scalar fields with (global) gauge symmetry group, $O(N), U(N)$, 
or $USp(2N)$, respectively. We shall sketch one step of the proof which goes 
back to \cite{NST02} (and corresponds to the argument in Sect. 5.7 of 
\cite{MZ}), choosing for the sake of definiteness the case of the $u(\infty, 
\infty)$ Lie algebra generated by the commutation relations (CR) (\ref{CRW}). 
(We refer for detail to \cite{BNRT} (a).)

The conformal Lie algebra in D dimensions ${\mathcal C}=so(D,2)$ can be viewed 
as a Lie subalgebra of the suitable centrally extended completion ${\hat u}$ of
 the infinite Lie algebra $u(\infty, \infty)$. The maximal compact subgroup 
$SO(2)\times SO(D)$ of the conformal group has a one-dimensional centre $SO(2)$.
 Its (hermitean) generator $H$ the {\it conformal Hamiltonian} is positive 
definite whenever the standard Minkowski space Hamiltonian $P_0$ is. Furthermore, it belongs to a compact Cartan subalgebra of ${\hat u}$ (contained in a completion of $u(\infty)\oplus u(\infty)$) has a (finitely degenerated) discrete spectrum in any unitary positive energy irreducible representation (UPEIR) of ${\hat u}$. We consider a basis $\{f_i\}$ of eigenfunctions of $H$. (Such a basis is constructed in Appendix A to \cite{BNRT} (a) in terms of homogeneous harmonic polynomials in a complex realization of the conformal compactification of Minkowski space.) A local field smeared with $f_i$ behaves as a creation operator while smearing with the conjugate ${\bar f}_i$ gives rise to an annihilation operator. We consider the n-th order determinant of mutually commuting (double) annihilation operators
\begin{equation}
\label{Dn}
D_n (= D_n^{(r)}) = det(X_{ij})|_{i,j=1}^n, \, X_{ij} = W({\bar f}_{r+i}, {\bar f}_{r+j})
\end{equation}
where $W(f, g)$ is the smeared bilocal field $W$ (and the superscript $r$ is designed to indicate that we are not necessarily starting with the lowest energy eigenvalue - a possibility, actually used in the argument of Sect. 2.2 of \cite{BNRT} (a)). A groundstate $|{\bf h}>$ of an UPEIR of ${\hat u}$ is a minimal energy eigenstate of the Cartan subalgebra that is annihilated by all lowering operators including $X_{ij}$. The norm square of the vector $D_n^*|{\bf h}>$ can be 
computed algebraically (using the CR of ${\hat u}$) and is a polynomial $p_n(N)$
 of degree $n$ in the central charge $N$ defined in (\ref{CRW}). In order to 
determine this polynomial we shall exploit our knowledge of its behaviour for 
integer values of $N$. In fact, in a realization of $W$ as a sum of normal 
products of free complex massless scalar fields $N$ is an integer, 
\begin{equation}
\label{mfr}
W(x, y) = \sum_{i=1}^m :\varphi_i^*(x) \varphi_i(y): \Rightarrow N=m.
\end{equation}
For such a $W$ $p_n(N)$ will vanish if $m<n$ and will be positive for $m=n$. 
This implies that for general $N$
\begin{equation}
\label{pn}
p_n(N) = <{\bf h}|D_n D_n^*|{\bf h}> \sim N(N-1)...(N-n+1)
\end{equation}
with an irrelevant positive coefficient. The nonnegativity of $p_n(N)$ for all 
$n$ implies that $N$ must be a positive integer.  
Then one proves (see Theorem 1 of \cite{BNRT} (a)):

(i) {\it All UPEIRs of ${\hat u}$ are realized (with multiplicities) in the Fock space of $N$ free massless complex scalar fields by (\ref{mfr}) (with $m=N$).}

(ii) {\it The ground states of equivalent representations of ${\hat u}$ in the Fock space form irreducible representations (IRs) of the gauge group $U(N)$. This establishes a one-to-one correspondence between the IRs of ${\hat u}$ occurring in the Fock space and the IRs of $U(N)$.}

\smallskip

\section{Discussion}

Considering together our earlier work \cite{NT, NST02, NST03, NRT08, BNRT} on 
globally conformal invariant QFT in an even number of space-time dimensions and
 the recent paper of Maldacena and Zhiboedov \cite{MZ} on 3D CFT with higher 
spin symmetry, we came to the conclusion that the crucial property which 
generalizes chiral (meromorphic) 2D CFT to higher (both even and odd) dimensions
 is the presence of an infinite series of conserved tensor currents. It turns 
out that both the main result of \cite{MZ}, namely:
 
{\it Any $D=3$ CFT with a unique rank two conserved current $J_2$ (with the 
properties of the energy momentum tensor) which admits higher rank conserved 
currents $J_s, s>2$, is generated by a bilocal field that is a sum of normal 
products of free (scalar or spin $\frac{1}{2}$) fields.}

and the argument designed to prove it parallels (as shown in Sect. 4) our earlier results and argument, valid for a theory with higher spin symmetry generated by a scalar field of dimension $D-2$ in a D-dimensional space-time. 

It was noted back in \cite{NST03} that in a $D=4$ GCI QFT generated by a scalar field of dimension four with the properties of a gauge field Lagrangian the methods used earlier (in \cite{NST02}) do not imply such a no go result. Furthermore, the three point function (\ref{TF}) 
provides a structure absent for $D=3$, for which the above mentioned arguments 
do not seem to apply. Even more suggestive, the new 4-point function of 
the (spin one) $U(1)$-current, discovered by Stanev \cite{S},
\begin{eqnarray}
\label{4j}
<J^{\alpha_1}(x_1) J^{\alpha_2}(x_2) J^{\alpha_3}(x_3) J^{\alpha_4}(x_4)> =
\Pi_{i=1}^4 \epsilon^{\alpha_i \mu_i \nu_i \rho_i} \, \, \nonumber \\
R_{\mu_1 \mu_2}(x_{12})R_{\nu_1 \nu_3}(x_{13})R_{\rho_1 \rho_4}(x_{14})
R_{\rho_2 \rho_3}(x_{23})R_{\nu_2 \nu_4}(x_{24})R_{\mu_3 \mu_4}(x_{34}), 
\end{eqnarray}
which does not appear to have a free field realization, and the related 
3-point function (\ref{JJF}) seem to leave open the way for constructing a 
non-trivial rational CFT model in 4D.

\smallskip

{\bf Acknowledgments}. I thank Yassen Stanev for collaboration, Karl-Henning 
Rehren and Raymond Stora for their interest and critical remarks, and Ali 
Tavanfar for inviting me to present a talk on this topic to the TH Journal Club. 
The hospitality and support of the Theory Division of CERN is gratefully 
acknowledged. The author's work has been supported in part by grant DO 02-257 
of the Bulgarian National Science Foundation.

\bigskip


\renewcommand{\theequation}{A.\arabic{equation}}

\setcounter{equation}{0}

\section*{Appendix A. Conformal invariants in four and three dimensions}\label{secAA} 
{\bf A1. Multiple role of Pauli matrices}

We introduce two copies, $\sigma^\mu$ and ${\tilde \sigma}_\mu$ of the same two-by-two matrices, assuming that they transform under inequivalent representations of $SL(2, {\mathbb C})$:
\begin{equation}
\sigma^0(=-\sigma_0)={\bf 1}={\tilde \sigma}_0(=-{\tilde \sigma}^0), \sigma^i(=\sigma_i)={\tilde \sigma_i},
\end{equation}
equipped with different undotted and dotted indicies $A,{\dot B}=1, 2$:
\begin{equation}
\label{x/d}
{\tilde x}=x^\mu{\tilde \sigma}_\mu=({\tilde x}^{A{\dot B}}), \, \partial  = \partial_\mu \sigma^\mu =
(\partial_{{\dot A} B})
\end{equation}
and use the skew-symmetric tensor 
\begin{equation}
\label{eps}
(\epsilon^{AB})=({\bar \epsilon}^{{\dot A}{\dot B}})(=(\epsilon_{AB})), \, \epsilon^{12}=1
\end{equation}
for raising (and lowering) indices. The matrix valued vectors $\sigma^\mu$ and ${\tilde \sigma}_\nu$ obey the exchange relations
\begin{equation}
\label{exch}
{\tilde \sigma}_\mu \sigma^\nu + {\tilde \sigma}_\nu \sigma^\mu = 2\delta_\mu^\nu (\delta^A_B), \, \,
\sigma^\mu{\tilde \sigma}_\nu + \sigma_\nu {\tilde \sigma}^\mu = 2\delta_\mu^\nu (\delta_{\dot A}^{\dot B}).
\end{equation}
The generators of the two (complex conjugate to each other) 2-dimensional representations of 
$SL(2,{\mathbb C})$ can be written as 
\begin{equation}
\label{smn}
\sigma_{\mu \nu}=\frac{1}{2}({\tilde \sigma}_\mu \sigma_\nu - {\tilde \sigma}_\nu \sigma_\mu)=
{\tilde \sigma}_\mu \sigma_\nu - \eta_{\mu \nu} =
((\sigma_{\mu\nu})^A_B), \, {\tilde \sigma}_{\mu\nu}=\frac{1}{2}(\sigma_\mu {\tilde \sigma}_\nu -
\sigma_\nu {\tilde \sigma}_\mu)(=-\sigma_{\mu\nu}^*) 
\end{equation}
($(\eta_{\mu \nu})= diag(- + + +))$ and satisfy the (anti)selfduality condition
\begin{equation}
\label{star}
*\sigma_{\mu\nu}:=\frac{1}{2}\epsilon_{\mu\nu\kappa\lambda}\sigma^{\kappa\lambda} =i\sigma_{\mu\nu},
*{\tilde \sigma}_{\mu\nu}=-i{\tilde \sigma}_{\mu\nu} \, \, (\mbox{for} \, \, \epsilon^{0123}=-\epsilon_{0123}=1).
\end{equation}
The tensor valued matrices ({\ref{smn}) can be also used to decompose a (real) 
skew-symmetric tensor $F_{\mu \nu}$ (like the Maxwell field) into (complex) 
irreducible components $(1, 0) \oplus (0, 1)$ with respect to 
$SL(2, {\mathbb C})$:
\begin{eqnarray}
\label{FAB}
F_{\mu \nu} = F^A_B (\sigma_{\mu \nu})^B_A + c.c. \Rightarrow F^A_B = \frac{1}{4} F_{\mu \nu}
(\sigma^{\mu \nu})^A_B, \nonumber \\ F^S_S = 0, F^{AB}= F^A_S\epsilon^{SB} = F^{BA}, \,
{\bar \sigma}_{\mu\nu}{\bar \epsilon}={\bar \epsilon}{\tilde \sigma}_{\mu\nu}. 
\end{eqnarray}  

\smallskip
 
{\bf A2. Two and three point conformal invariants for $D=4$.}

The 2-component spinors $\lambda_1 , \lambda_2$ in (\ref{P12}) are translation invariant but transform non-trivially under the (Weyl) conformal inversion (that is a proper conformal transformation, - i.e., belongs to the connected component of $SU(2, 2)$), 
\begin{equation}
\label{Iw}
I_w: (x^0, {\bf x}) \rightarrow \frac{(x^0, -{\bf x})}{x^2} \, \, \, ({\check x}{\tilde \sigma}\rightarrow -x\sigma), \, \lambda_1\rightarrow \lambda_1{\check x}_1{\tilde \sigma}, {\bar \lambda}_2\rightarrow
{\check x}_2{\tilde \sigma}{\bar \lambda}_2. 
\end{equation}
(We note that while $I_w$ acts on the 4-vector $x$ as an involution, the square of its action on $\lambda$ gives $-\lambda$, ${\bf -1}$ being the non-trivial central element of $SL(2,{\mathbb C})$
which is mapped on the identity of the (connected) Lorentz group $SO_0(3, 1)$.)  It is a simple exercise to check that $P_{12}$ is invariant under $I_w$. In order to verify the formula (\ref{PP}) for $P_{12}P_{21}$ one uses
\begin{eqnarray}
\label{PPR} 
\zeta_{i\mu}=\lambda_i{\tilde \sigma}_\mu{\bar \lambda}_i \Rightarrow \zeta_{i\mu}(\sigma^\mu)_{{\dot A}B}=
2{\bar \lambda}_{i{\dot A}}\lambda_{iB}, \, i=1, 2, \, \zeta_{12}^\mu:=\lambda_1{\tilde \sigma}^\mu{\bar \lambda}_2 = {\bar \zeta}_{21}^\mu, \, \nonumber \\
\eta^{\mu\nu}{\tilde \sigma}_\mu^{A_1{\dot B}_1}\otimes{\tilde \sigma}_\nu^{A_2{\dot B}_2} = -2 \epsilon^{A_1A_2} {\bar \epsilon}^{{\dot B}_1{\dot B}_2}, \, {\tilde \sigma}^\mu\sigma_\nu{\tilde \sigma}_\mu = -2 {\tilde \sigma}_\nu \nonumber \\
\Rightarrow \zeta_{12}\zeta_{21} =-\zeta_1\zeta_2, \zeta_{12}\zeta_i = 0, \, i=1, 2.
\end{eqnarray}
The derivation of the cubic relation (\ref{PPP}) requires more work. One again applies (\ref{PPR}) to write 
\begin{eqnarray}
\label{Ptr}
2x_{12}^2 x_{23}^2 x_{13}^2(P_{12}P_{23}P_{31} - P_{32}P_{21}P_{13}) =  \nonumber \\
-\frac{1}{4}tr[(\zeta_1\sigma {\tilde x}_{12} \zeta_2\sigma {\tilde x}_{23} \zeta_3\sigma - \zeta_3\sigma {\tilde x}_{23}\zeta_2\sigma {\tilde x}_{12} \zeta_1 \sigma){\tilde x}_{13}],
\end{eqnarray}
then uses $x_{13}=x_{12}+x_{23}$ and ${\tilde x} \zeta\sigma {\tilde x} = 2(x\zeta)x - x^2 \zeta$
to reduce the problem to the trace of the product of four two-by-two matrices where one finally applies the 
formula
\begin{eqnarray}
\label{tr4}
\frac{1}{2}tr(a{\tilde b}c{\tilde d}) = ab \, cd -ac \, bd +ad \, bc +i\epsilon^{\kappa \lambda \mu \nu} a_\kappa b_\lambda c_\mu d_\nu \, \, (\epsilon^{0123} = 1) \nonumber \\  \Rightarrow
\frac{1}{4}tr(a{\tilde b}c{\tilde d +c{\tilde b}a\tilde d}) = ab \, cd -ac \, bd +ad \, bc  \nonumber \\
\mbox{for} \, \, a= a_\mu \sigma^\mu, \, {\tilde b} = b^\mu{\tilde \sigma}_\mu, \, ab = a_\mu b^\mu.
\end{eqnarray}

Both the two- and the three-point invariants, $P_{12}$ and $L_i$  are homogeneous of degree 
$(1, 1)$ in $(\lambda_i, {\bar \lambda}_j)$. We now proceed to displaying 3-point invariants of degree $(3, 1)$ (and their conjugate of degree $(1, 3)$) which, clearly, cannot be expressed in 
terms of the preceding. To this end we first introduce a skew-symmetric tensor (\ref{Fomeg}) of degree $(2, 0)$
\begin{equation}
\label{omega}
\omega_{\mu \nu}(=\omega(\lambda,\lambda)_{\mu \nu}) = \lambda \sigma_{\mu \nu} \epsilon \lambda. 
\end{equation}
For fixed indices $\omega$ defines a quadratic form in $\lambda$, which gives rise (by a standard polarization procedure) to a symmetric bilinear form $\omega(\lambda_1, \lambda_2)$. Setting $\omega(i):=\omega(\lambda_i, \lambda_i)$ and using (\ref{PPR}) we can write
\begin{equation}
\label{o1o2}
\omega(1)_{\mu \rho} \eta^{\rho \tau}\omega(2)_{\tau \nu} = \omega(\lambda_1, \lambda_2)_{\mu \nu} \, \lambda_1\epsilon\lambda_2.
\end{equation}
It follows that for $\lambda_1=\lambda_2$ the product (\ref{o1o2}) vanishes.
In view of the self-duality condition (\ref{star}) $\omega_{\mu \nu}$ has only three linearly independent (complex) components. It is convenient to choose a basis of light-cone and transverse projections (which are real for real $\lambda$ - cf. (\ref{x3D})):
\begin{equation}
\label{opm}
\omega_+ := \frac{1}{2}(\omega_{13} + \omega _{10}) = \lambda_1^2, \, \omega_-:= \frac{1}{2}(\omega_{13} - \omega_{10}) = \lambda_2^2, \, \omega_0:= \frac{1}{2}\omega_{03} = \lambda_1 \lambda_2. 
\end{equation}  
The inner product of two such vectors ${\bf \omega}(i)= 
{\bf \omega}(\lambda_i), i=1, 2$, is given by
\begin{eqnarray}
\label{inprod}
{\bf \omega}(1){\bf \omega}(2) := \frac{1}{8}\omega^{\mu \nu}(1)\omega_{\mu \nu}(2) = \nonumber \\ 
\omega_+(1)\omega_-(2)+
\omega_-(1)\omega_+(2)+2\omega_0(1)\omega_0(2)=(\lambda_{11} \lambda_{22}-\lambda_{12} \lambda_{21})^2.
\end{eqnarray}
In particular, we see that the 3-vector ${\bf \omega}={\bf \omega}(\lambda)$ is isotropic:
\begin{equation}
\label{null}
{\bf \omega}^2 = 2(\omega_+ \omega_- - \omega_0^2) = 0
\end{equation}  
(something that also follows from (\ref{o1o2})). Consider now the 2- and 3-point 4-vectors $R_{i3}^\mu, i=1, 2$ and $L_3^\mu$ (with $\zeta_3$ replaced by the 4-vector index $\mu$:
\begin{equation}
\label{RLmu}
R_{i3}^\mu = {\check x}_{i3}^2 \zeta_i^\mu - 2(\zeta_i {\check x}_{i3}) \check x_{i3}^\mu, \,
i=1, 2; \, \, L_3^\mu = {\check x}_{23}^\mu - {\check x}_{13}^\mu. 
\end{equation}
The self-dual projections of their skew-symmetric products,
\begin{equation}
\label{RoL}
RL_{i3}:= \frac{1}{2}R_{i3}^\mu \omega_{\mu \nu} L_3^\nu, \, i = 1, 2, \, \omega=\omega(3)
\end{equation}
are conformal invariant. In verifying this we observe that under the Weyl inversion (\ref{Iw})
$\omega$ transforms as
\begin{eqnarray}
\label{Iomeg}
(I_w \omega)_a = (x^2)^{-2} V_a^b \omega_b, \, a, b = +, -, 0, \, x = x_3, \nonumber \\
V_+^+ = (x^+)^2, V_+^- = {\bar y}^2, V_+^0= 2x^+ {\bar y}, \, V_-^+ = y^2, V_-^-=(x^-)^2,
\nonumber \\ V_-^0 = 2 y x^-, \, V_0^+ = y x^+, V_0^- = {\bar y} x^-, V_0^0=x^+x^- + y{\bar y}.  
\end{eqnarray}
This transformation law satisfies, as expected,
\begin{equation}
\label{Iomsq}
(I_w {\bf \omega})^2 = (x^2)^{-2}{\bf \omega}^2 = 0.
\end{equation}

\smallskip

{\bf A3. The fermionic box diagram: a four point invariant}

The following relation, computed by Yassen Stanev, demonstrates that the real 
part of the product of four P-invariants can be only expressed (in a rather 
complicated way) in terms of R- and L-invariants on the price of introducing 
fake singularities (in $x_{13}^2$ and $x_{24}^2$):       
\begin{eqnarray}
\label{4P}
4(P_{12}P_{23}P_{34}P_{41} + P_{14}P_{43}P_{32}P_{21}) = (R_{12}+2 L^1_{24}L^2_{31})
(R_{34} + 2 L^3_{42}L^4_{13}) + \, \, \, \nonumber \\
 (R_{14}+2L^1_{24}L^4_{13})(R_{23}+2L^2_{31}L^3_{42}) - 4L^1_{24}L^2_{31}L^3_{42} 
L^4_{13} \, \, \, \nonumber \\
-\frac{1}{t}(R_{12}+ 2 L^1_{23}L^2_{41})(R_{34}+ 2 L^3_{41}L^4_{23}) 
-\frac{1}{s}(R_{14}+ 2 L^1_{34}L^4_{12})(R_{23}+ 2 L^2_{34}L^3_{12}) \, \, \, 
\nonumber \\
+ \frac{2}{t}(L^1_{23}L^3_{41}R_{24}+R_{13}L^2_{41}L^4_{23}) +
\frac{2}{s}(L^1_{34}L^3_{12}R_{24}+R_{13}L^2_{34}L^4_{12}) \, \, \, \nonumber \\
+\frac{s}{t}R_{12}R_{34} +\frac{t}{s}R_{14}R_{23} +\frac{1-s-t}{st}R_{13}R_{24}, \,
 \, \, \, \,  \, \, \,
\end{eqnarray}
where $s, t$ are the independent cross-ratios
\begin{equation}
\label{st}
s=\frac{x_{12}^2 x_{34}^2}{x_{13}^2 x_{24}^2}, \, t=\frac{x_{14}^2 x_{23}^2}
{x_{13}^2 x_{24}^2}. \, \,
\end{equation}
The expression in the right hand side of (\ref{4P}) is not unique because of 
relations of the type
\begin{equation}
\label{Lrel}
L^4_{12} + L^4_{23} = L^4_{13}. 
\end{equation}
In any case, it appears advantageous to use $P_{ij}$ and a set of independent
$L^k_{ij}$ as a basis of confornmal invariant spin-tensors of dimension $(1,1)$
in $(\lambda, {\bar \lambda})$ even when they may be expfressed in  terms of 
$R_{ij}$ and $L^k_{ij}$. (One has to add to this list invariants of 
higher weight like $RL_{i3}$ (\ref{RoL}) and their conjugates.) 
 
\smallskip

{\bf A4.  Reduction to the D=3 case.}

In three dimension the 4D spinorial Lorentz group $SL(2, {\mathbb C})$ reduces to its real subgroup $SL(2, {\mathbb R})$ so that the spinors $\lambda_i$ can be chosen real. We then only need one set of {\it symmetric Pauli matrices}\footnote{The matrices $\sigma_0 = {\bf 1}, \sigma_1, \sigma_3$ actually correspond to $({\tilde \sigma}_\mu)$ but we drop the tilde sign.} $(\sigma_\pm, \sigma_1$ where $\sigma_\pm$ correspond to the light-cone coordinate $x^\pm$ (cf. (\ref{x3D})): $x^0 + x^3 \sigma_3 = x^+\sigma_+ + x^-\sigma_-$. We shall feel free to use the orthogonal or the lightcone (and transverse) coordinates depending on convenience, taking for the line element either of the two equivalent forms:
\begin{equation}
\label{pm}
dx^2 = (dx^1)^2 +(dx^3)^2 - (dx^0)^2 = dy^2 -dx^+dx^-. 
\end{equation}
The exchange relations (\ref{exch}) assume a familiar form in terms of the generators $\gamma_\mu$ of the Clifford algebra $Cl(2, 1)$ of 3-dimensional Minkowski space (cf. \cite{T11}) which are two-by-two real traceless matrices:
\begin{equation}
\label{gam}
\gamma_\mu = \sigma_\mu \epsilon, \, \mu= 0, 1, 3 \, (\mbox{or} \, +, -, 1) \Rightarrow [\gamma^\mu, \gamma_\nu]_+ := \gamma^\mu \gamma_\nu + \gamma_\nu \gamma^\mu = 2\delta^\mu_\nu (\delta^A_B).
\end{equation}
The relations (\ref{PPR}) on the other hand assume the form
\begin{equation}
\label{gog}
\gamma^{\mu A_1}_{B_1}\gamma^{A_2}_{\mu B_2} = \delta^{A_1}_{B_2}\delta^{A_2}_{B_1} +
\epsilon^{A_1A_2}\epsilon_{B_2B_1};
\end{equation}
\begin{eqnarray}
\label{zet}
\zeta_i^\mu=\lambda_i\sigma^\mu\lambda_i, \, i=1, 2, \, \zeta_{12}^\mu = \lambda_1\sigma^\mu \lambda_2=\zeta_{21}^\mu \, \Rightarrow \nonumber \\
\zeta_{12}^2 = (\lambda_1\epsilon\lambda_2)^2 =-\frac{1}{2}\zeta_1\zeta_2, \, \zeta_i^2 =0=\zeta_{12}\zeta_i. 
\end{eqnarray}



\smallskip


\end{document}